\documentclass{webofc}
\usepackage[varg]{txfonts}   % Web of Conferences font
\usepackage{siunitx}
\usepackage[super]{nth}
\begin{document}
\title{Muonium Lamb shift: theory update and experimental prospects}
\author{\firstname{Gianluca} \lastname{Janka}\inst{1}\and
\firstname{Ben} \lastname{Ohayon}\inst{1} \and
\firstname{Paolo} \lastname{Crivelli}\inst{1}\fnsep\thanks{\email{crivelli@phys.ethz.ch}}
        % etc.
}

\institute{ETH Zurich, Institute for Particle Physics and Astrophysics, 8093 Zurich, Switzerland}

\abstract{%
 We review the theory of the Lamb shift for muonium, provide an updated numerical value and present the prospects of the Mu-MASS collaboration at PSI to improve upon their recent measurement. Due to its smaller nuclear mass, the contributions of the higher-order recoil corrections (\SI{160}{\kilo\hertz} level) and nucleus self-energy (\SI{40}{\kilo\hertz} level) are enhanced for muonium compared to hydrogen where those are below the level of the latest measurement performed by Hessels et al. and thus could not be tested yet.
The ongoing upgrades to the Mu-MASS setup will open up the possibility to probe these contributions and improve the sensitivity of this measurement to searches for new physics in the muonic sector.
}
\maketitle
\section{Introduction}
\label{intro}
Over the years, the $2S_{1/2}-2P_{1/2}$ Lamb shift (LS) determination in hydrogen (H) has improved to the point that it became sensitive to the finite size effects of the nucleus \cite{1981_Lundeen}. To use the LS measurements to probe bound-state QED required an independent and accurate determination of the proton charge radius. Prior to \num{2010}, the values for the proton charge radius determined from e-p scattering experiments and from hydrogen spectroscopy were in agreement \cite{2014_CODATA}. With a new method based on muonic hydrogen laser spectroscopy, the proton charge radius was extracted with more than \num{10} times precision\cite{2010_Pohl, 2013_Antognini}. Unexpectedly, the radius was found to be around \SI{4}{\percent} smaller than from the other measurements, which resulted in a disagreement of about $7\sigma$. This led to what was dubbed as the proton charge radius puzzle \cite{2013_Pohl}, triggering many experimental and theoretical efforts striving for a solution. The most recent determinations of the radius from e-p scattering \cite{2019_Xiong} or from the Lamb shift \cite{2019_Hessels} and the $2S-4P$ transition \cite{2017_Beyer} in hydrogen are agreeing with the value from muonic hydrogen, while the most recent measurement of the $2S_{1/2} - 8D_{5/2}$ transition in hydrogen is disagreeing by around $3\sigma$ \cite{2021_Brandt}. Measurements of the $1S-3S$ transition in hydrogen support both the smaller radius \cite{2020_Grinin} as well as the originally accepted value \cite{2018_Fleurbaey}.  Although the puzzle is arguably coming to an end, the discrepancy between the experiments is yet to be understood \cite{2020_Karr}.

Muonium (M), which is composed of a positively charged muon orbited by an electron, is an excellent candidate to probe bound-state QED due to its lack of internal structure and therefore freeing it from finite size effects. Furthermore, M is lighter than hydrogen, making it more suitable for testing effects related to its nucleus mass such as the recoil or the nucleus self-energy. The last numerical value of the M LS calculated with bound-state QED was by J.~R. Sapirstein and D.~R. Yennie \cite{1990_Sapirstein} to be at \num{1047.49}($1$)($9$) \si{\mega\hertz} in $\num{1990}$. Here we provide an updated value which takes into account the improvements in the LS calculations done in the last three decades. 

\section{Lamb shift calculation}
\label{sec-1}

To calculate the energy difference between the $2S_{1/2}$ and $2P_{1/2}$ states, we followed the theory of the hydrogen atom as summarized by CODATA 2018 \cite{2021_CODATA} and added some relevant terms for muonium which due to their smallness in hydrogen were neglected. In the following, we use $m_e$ for the electron, $m_n$ for the nuclear and $m_r$ for the reduced mass. The reduced mass is defined as:
\begin{equation}
m_r = \frac{m_e m_n}{m_e+m_n}
\end{equation}

We first tabulate the non-recoil contributions, which are identical between muonium and hydrogen up to powers of $\left(\frac{m_r}{m_e}\right)$. The main contribution to the Lamb shift is the self-energy $E_\text{SE}$, where the bound electron emits and reabsorbs a virtual photon. A similar effect comes from the vacuum polarization, which can be looked at as an effect from the self-energy of the gauge boson, where it produces a virtual e$^+$e$^-$ ($E_\text{VP}$), or in more suppressed cases $\mu^+\mu^-$ ($E_{\text{VP}\mu}$) or hadronic ($E_\text{VP,had}$) pair, which eventually annihilate again. The contributions can be calculated as:
\begin{align}
%% SE %%
E_\text{SE} &=\frac{\alpha}{\pi}\frac{(Z\alpha )^4}{n^3}
\left(\frac{m_r}{m_e}\right)^3
F(Z\alpha) ~m_e c^2 \\
%% VP %%
E_\text{VP}  &=\frac{\alpha}{\pi}\frac{(Z\alpha)^4}{n^3}\left(\frac{m_r}{m_e}\right)^3H(Z\alpha)~m_e c^2 \\
%% VPmuhad %%
E_{\text{VP}\mu} &=
\frac{\alpha}{\pi}\frac{ (Z\alpha )^4}{n^3} 
\left(\frac{m_r}{m_e}\right)^3
\left(\frac{m_e}{m_\mu}\right)^2
\left[-\frac{4}{15}\right]
\delta_{l0} 
~m_ec^2\\
E_\text{VP,had} &=0.671(15)~E_{\text{VP}\mu} \\
\nonumber\\ 
F(Z\alpha) &= L~A_{41} + A_{40} + (Z\alpha)A_{50} +(Z\alpha)^2\left[L^2A_{62}+L~A_{61}+G_\text{SE}(Z\alpha)\right]\\
H(Z\alpha) & = V_{40} + (Z\alpha)V_{50} + (Z\alpha)^2\left[L~ V_{61} + G_\text{VP}(Z\alpha) + G_{\text{VP}}^{(R)}(Z\alpha)\right],
\end{align}
where we denote $L$ = ln~($\frac{m_e}{m_r}\frac{1}{(Z\alpha)^2}$) and $\delta_{ij}$ as the Kronecker delta.

Higher-order nonrecoil contributions result from two- and three-photon corrections $E_\text{2ph}$ and $E_\text{3ph}$: 
\begin{align}
E_\text{2ph} &=\left(\frac{\alpha }{\pi }\right)^2\frac{ (Z\alpha)^4}{n^3}\left(\frac{m_r}{m_e}\right)^3 F_\text{2ph}\left(Z\alpha\right)m_e c^2\\
E_\text{3ph}&=\left(\frac{\alpha }{\pi }\right)^3\frac{ (Z\alpha)^4}{n^3}\left(\frac{m_r}{m_e}\right)^3F_\text{3ph}\left(Z\alpha\right)m_e c^2\\    
\nonumber\\
\nonumber F_\text{2ph}(Z\alpha) &= B_{40} + (Z\alpha) B_{50} + (Z\alpha)^2\left[L^3B_{63} + L^2B_{62} + L~B_{61} + B_{60}\right]\\
  &~~~+(Z\alpha)^3\left[L^2B_{72} + L~B_{71} + ... \right] \\
F_\text{3ph}(Z\alpha) &= C_{40} + (Z\alpha) C_{50} + (Z\alpha)^2\left[L^3 C_{63}+ L^2C_{62} + L~C_{61} + C_{60} + ...\right].
\end{align}
\newpage
Compared to hydrogen, the nucleus of muonium is approximately nine times lighter, making it more sensitive to recoil corrections. We list the recoil corrections appearing in CODATA \num{2018} such as the Barker-Glover correction $E_\text{BKG}$, the relativistic recoil $E_\text{rec,S} + E_\text{rec,R}$, the radiative recoil $E_\text{RR}$ and the nucleus self-energy $E_\text{SEN}$ and later additional contributions potentially important to M:
\begin{align}
%% BK %%
E_\text{BKG} &= 
\frac{(Z\alpha )^4}{2 n^3}
\left(\frac{m_r}{m_e}\right)^3
\left(\frac{m_e}{m_n}\right)^2
\frac{1-\delta_{l0}}{\kappa~(2 l+1)}
m_ec^2\\
%% Rec,S%%
\label{eq:recs}
E_\text{rec,S} &= 
\frac{(Z\alpha )^5}{ \pi  n^3}
\left(\frac{m_r}{m_e}\right)^3
\left(\frac{m_e}{m_n}\right)~
S(Z\alpha)~ m_e c^2\\
%% Rec,R%%
\label{erec_r}E_\text{rec,R} &=
\frac{ (Z\alpha )^6}{n^3}
\left(\frac{m_r}{m_e}\right)^3
\left(\frac{m_e}{m_n}\right)~
R(Z\alpha)
~m_ec^2 \\
%% RR %%
\label{err} E_\text{RR} &= 
\frac{\alpha}{\pi}
\frac{  (Z \alpha )^5}{\pi  n^3}
\left(\frac{m_r}{m_e}\right)^3
\left(\frac{m_e}{m_n}\right)~
Q(Z\alpha)~
\delta_{l0}~
m_e c^2 \\
%% SEN %%
\label{eq:esen}E_\text{SEN} &=
\frac{ Z^2 \alpha}{\pi}
\frac{ (Z \alpha )^4}{n^3} 
 \left(\frac{m_r}{m_e}\right)^3
\left(\frac{m_e}{m_n}\right)^2
\left[\left(\boldsymbol{\frac{10}{9}}+\frac{4}{3}\text{ln}\left[\frac{m_n}{m_r(Z \alpha )^2}\right]\right)\delta_{l0}-\frac{4}{3}\text{lnk$_0$}\right]
m_e c^2\\
 %% Functions %%
\nonumber \\
\label{eq:recs_s}
S(Z\alpha) & = L~D_{51} + D_{50} -\frac{2\delta_{l0}}{m_n^2-m_e^2}
    \left[m_n^{2}\text{ln}\left(\frac{m_e}{m_r}\right)
    -m_e^{2}\text{ln}\left(\frac{m_n}{m_r}\right)\right] \\
R(Z\alpha) &= D_{60}  + (Z\alpha)~G_\text{REC}(Z\alpha) \\
\label{eq:rr_q}
Q(Z\alpha) & = 6 \zeta[3]-2 \pi ^2\text{ln}\left(2\right)+\frac{35 \pi ^2}{36}-\frac{448}{27}+ (Z \alpha )\left[\frac{2}{3}\pi~\text{ln}\left((Z \alpha )^{-2}\right)^2\right] + ... 
\end{align}
where $\kappa$ is the angular momentum parity quantum number, $\zeta$ is the Riemann zeta function and lnk$_0$ is the relevant Bethe logarithm. In atoms with a nuclear size such as hydrogen, the $\frac{10}{9}$ term in E$_\text{SEN}$ (Eq.~\ref{eq:esen}) would be absorbed in the finite size contributions. Since muonium has no sub-structure, it is free from finite size effects and therefore the finite nuclear size contributions can be omitted but $E_\text{SEN}$ has to be extended. See also the discussion in Ref.~\cite{1995_Pachucki}.

All the coefficients $A$, $B$, $C$, $D$ and $V$, the remainder functions $G$ as well as the uncertainties of each contribution are tabulated in Ref.~\cite{2021_CODATA, 2019_Pachucki, 2019-Saveli, 2019-Saveli2}. These quantities typically depend on the quantum numbers $n$ (principal), $l$ (azimuthal) and $j$ (total angular momentum). Their notation is suppressed here for brevity.

%%% Higher order recoil terms %%
In addition to the above, we collect higher-orders recoil contributions. These are negligible for H, but are up to two orders of magnitude larger for M, and, thus, should be considered. We first focus on radiative-recoil contributions. Czarnecki and Melnikov \cite{2002-Czar} have calculated higher-order mass contributions to the radiative insertions in the electron line, of which the \nth{2} order is
\begin{align}
E_\text{RR2e} &= 
\frac{\alpha}{\pi}
\frac{  (Z \alpha )^5}{ n^3}
\left(\frac{m_r}{m_e}\right)^3
\left(\frac{m_e}{m_n}\right)^2~
\left[\pi8\ln(2)-\pi\frac{127}{32}\right]~
\delta_{l0}~
m_e c^2 ,
\end{align}
which returns a negligible \SI{80}{\hertz} for the M LS.
Another radiative recoil contribution comes from polarization insertions calculated in \cite{1995-Eides}, whose \nth{2} order in mass correction is given in \cite{2021-Eides}
\begin{align}
E_\text{RR2p} &= 
\frac{\alpha}{\pi}
\frac{  (Z \alpha )^5}{ n^3}
\left(\frac{m_r}{m_e}\right)^3
\left(\frac{m_e}{m_n}\right)^2~
\left[-\frac{3\pi}{16}\right]~
\delta_{l0}~
m_e c^2 ,
\end{align}
returning a negligible \SI{-10}{\hertz} for the M LS. Very recently, spin-independent three-loop radiative corrections have been calculated for positronium and muonium \cite{2021-Eides}, giving
\begin{align}
E_\text{RR3} &= 
\left(\frac{\alpha}{\pi}\right)^2
\frac{  (Z \alpha )^5}{\pi n^3}
\left(\frac{m_r}{m_e}\right)^3
\left(\frac{m_e}{m_n}\right)~
\left[-11.4...\right]~
\delta_{l0}~
m_e c^2 ,
\end{align}
which returns a negligible \SI{-30}{\hertz} contribution to the M LS.

For pure recoil contributions, equations \ref{eq:recs} and \ref{eq:recs_s} account for recoil terms up to order $(Z\alpha)^5(m_e/m_n)^3\ln{(m_e/m_n)}$. The expansion in mass ratio of the pure recoil term of order $(Z\alpha)^6$ is calculated partially in \cite{2002-Czar}, which to \nth{2} order gives
\begin{align}
E_\text{rec,R2} &=
\frac{ (Z\alpha )^6}{n^3}
\left(\frac{m_r}{m_e}\right)^3
\left(\frac{m_e}{m_n}\right)^2
\left[
\frac{4} {\pi^2}\ln{\frac{m_n}{m_e}}-
\frac{8} {3}    \ln{\frac{m_n}{m_e}}-
\frac{12\zeta_3}{\pi^2}+   
\frac{3}{\pi^2}+
\frac{8}{3}
\right]
\delta_{l0}~
m_ec^2,
\end{align}
returning \SI{-575}{\hertz} for the M LS. This contribution is not negligible compared with the \SI{0.8}{\kilo\hertz} uncertainty attached to equation \ref{err} and \ref{eq:rr_q}. To account for it being partial, we take the full contribution as its uncertainty. The total uncertainty in the muonium lamb shift is thus \SI{1.0}{\kilo\hertz}.

%%% HFS %%%
As another addition to CODATA \num{2018}, we include the off-diagonal hyperfine-structure contribution $E_\text{HFS}$. This additional recoil contribution arises from the mixing of the fine structure sublevels due to the hyperfine interaction as given by for example V. Yerokhin et al. \cite{2019_Pachucki}:
\begin{equation}
    E_\text{HFS} =
    \frac{\alpha ^2 (Z \alpha)^2}{n^3}
    \left(\frac{m_e}{m_p}\right)^2
    \left(\frac{\mu}{\mu_\text{N}}\right)^2
    \frac{ 2 I(I+1)}{81}(-1)^{j+1/2}
    \delta_{l1} m_e c^{2}.
\end{equation}
The $I$ is the nuclear spin quantum number, $\mu$ the nuclear magnetic moment and $\mu_\text{N}$ the nuclear magneton.

For both hydrogen and muonium, we evaluate these contributions for the $2S_{1/2}$ and $2P_{1/2}$ levels. Their difference in frequency are summarized in Tab.~\ref{tab:calculations_summary}. We calculate an updated LS frequency for muonium while validating our result with the values summarized by Yerokhin for hydrogen \cite{2019_Pachucki}. The finite-size and other nuclear effects for hydrogen are not shown, but would amount to around $\SI{138}{\kilo\hertz}$. Only uncertainties $\geq$ \SI{500}{\hertz} are given. Our final value for the LS in M is \SI{1047.498+-0.001}{\mega\hertz}. It is in agreement and \num{100} times more accurate than the latest bound-state QED calculation by Sapirstein and Yennie \cite{1990_Sapirstein}. However, C. Frugiuele et al.~recently calculated the M LS using effective field theory to be \SI{1047.284+-0.002}{\mega\hertz} \cite{2019_Peset}, which differs around \SI{200}{\kilo\hertz} from our value and, thus, deserves further investigations.

Comparing the contributions for hydrogen and muonium, we identify the Barker-Glover correction as well as the nucleus self-energy as the two most interesting contributions for future experiments. The current best determination of the hydrogen Lamb shift has an uncertainty of around \SI{3}{\kilo\hertz} \cite{2019_Hessels}, hence these two contributions could not be resolved so far. In muonium, they are enhanced and can be probed by achieving Lamb shift uncertainties below \SI{160}{\kilo\hertz} and \SI{40}{\kilo\hertz}, respectively. 

\begin{table}[b!h!t]

  \centering
  \caption[]{Summary of the calculated contributions to the hydrogen and muonium Lamb shift transition. The theoretical calculations from Sapirstein $\&$ Yennie \cite{1990_Sapirstein} and from Frugiuele et al. \cite{2019_Peset} are included for comparison. Uncertainties smaller than \SI{0.5}{\kilo\hertz} are not tabulated.}
    \begin{tabular}{l|lrr}
    \hline \noalign{\vspace{-1pt}} \hline
        \noalign{\smallskip}
                & Largest Order & Hydrogen & Muonium\hphantom{(0)} \\
                & & (MHz) & (MHz)\hphantom{(0)} \\
        \noalign{\smallskip}
         \hline
        \noalign{\smallskip}

~ $E_\text{SE}$  & $\alpha~(Z\alpha)^4~L$ & $1084.128 $ & $1070.940\hphantom{(0)}$ \\\\
~ $E_\text{VP}$  & $\alpha~(Z\alpha)^4$ & $ -26.853 $ & $-26.510\hphantom{(0)}$ \\
~ $E_{\text{VP}\mu+\text{had}}$ & $\alpha~(Z\alpha)^4(m_e/m_\mu)^2$ & $ -0.001$ & $-0.001\hphantom{(0)}$ \\\\

~ $E_\text{2ph}$ & $\alpha^2(Z\alpha)^4$ & $ 0.065$ & $0.065\hphantom{(0)}$ \\
~ $E_\text{3ph}$ & $\alpha^3(Z\alpha)^4$  & $ 0.000$ & $0.000\hphantom{(0)}$ \\\\

~ $E_\text{BKG}$    & $ ~ ~ ~ (Z\alpha)^4~ ~ ~ ~ (m_e/m_n)^{2}$  & $-0.002 $ & $-0.168\hphantom{(0)}$ \\
~ $E_\text{rec,S}$  & $ ~ ~ ~ (Z\alpha)^5~L~(m_e/m_n)$  & $ 0.358$ & $3.138\hphantom{(0)}$ \\
~ $E_\text{rec,R}$  & $ ~ ~ ~ (Z\alpha)^6~ ~ ~ ~ (m_e/m_n)$  & $ -0.001$ & $-0.012\hphantom{(0)}$\\
~ $E_\text{rec,R2}$ & $ ~ ~ ~ (Z\alpha)^6~ ~ ~ ~ (m_e/m_n)^2$  & $-0.000$ & $-0.001(1)$ \\
~ $E_\text{RR}$     & $\alpha~(Z\alpha)^5~ ~ ~~  (m_e/m_n)$  & $-0.002$ & $-0.014(1)$ \\
~ $E_\text{RR2e+p}$ & $\alpha~(Z\alpha)^5~ ~ ~~  (m_e/m_n)^2$  & $0.000$ & $0.000\hphantom{(0)}$\\
~ $E_\text{RR3}$ & $\alpha^2(Z\alpha)^5~ ~ ~~  (m_e/m_n)$  & $-0.000$ & $-0.000\hphantom{(0)}$
\\\\

~ $E_\text{SEN}$ & $Z^2\alpha~(Z\alpha)^4~(m_e/m_\mu)^{2}$ & $0.001$ & $0.041\hphantom{(0)}$ \\\\
~ $E_\text{HFS}$ & $\alpha^2(Z\alpha)^2~(m_e/m_n)^{2}$ & $0.002$ & $0.019\hphantom{(0)}$ \\\\

    \noalign{\smallskip}
    \hline \noalign{\smallskip} \hline
    \noalign{\vspace{2pt}}
~ Sum   &  & & $1047.498(1)$ \\
~ Ref. \cite{1990_Sapirstein} & &  & $1047.49(9)\hphantom{0}$ \\
~ Ref. \cite{2019_Peset} & & & $1047.284(2)$ \\
        \noalign{\smallskip}
         \hline
\end{tabular}
\label{tab:calculations_summary}
\end{table}

\section{Mu-MASS experiment}
\label{sec-2}
The Mu-MASS experiment was proposed in \num{2018} \cite{2018_Crivelli} with the main goal to improve upon the $1S-2S$ transition in muonium. Key elements of the experiment are the unique low energy muon (LEM) beam line at PSI, the efficient tagging of the incoming muons, the production of M at cryogenic temperatures \cite{2012_Antognini},  a high-power CW UV-laser for excitation \cite{2021_Burkley}  as well as an efficient detection of the metastable M(2S) atoms. To test the detection scheme, an intense M(2S) beam of the order of \SI{100}{\hertz} was formed, opening up the possibility to improve upon the muonium Lamb shift \cite{2020_Janka}.

The metastable beam is produced at the LEM beamline by guiding a continuous \SI{10}{\kilo\electronvolt} $\mu^+$ beam on a thin carbon foil ($\approx$ \SI{10}{\nano\meter}). While passing through the foil, the $\mu^+$ can pick up an electron to form M, predominately in the ground state, but also \SIrange{5}{10}{\percent} in the excited $2S$ state. During this process, the $\mu^+$ also releases secondary electrons, which are detected and used for tagging. 

\begin{figure}[tb!]
    \centering
    \includegraphics[width=\textwidth,trim={0 0 0 0},clip]{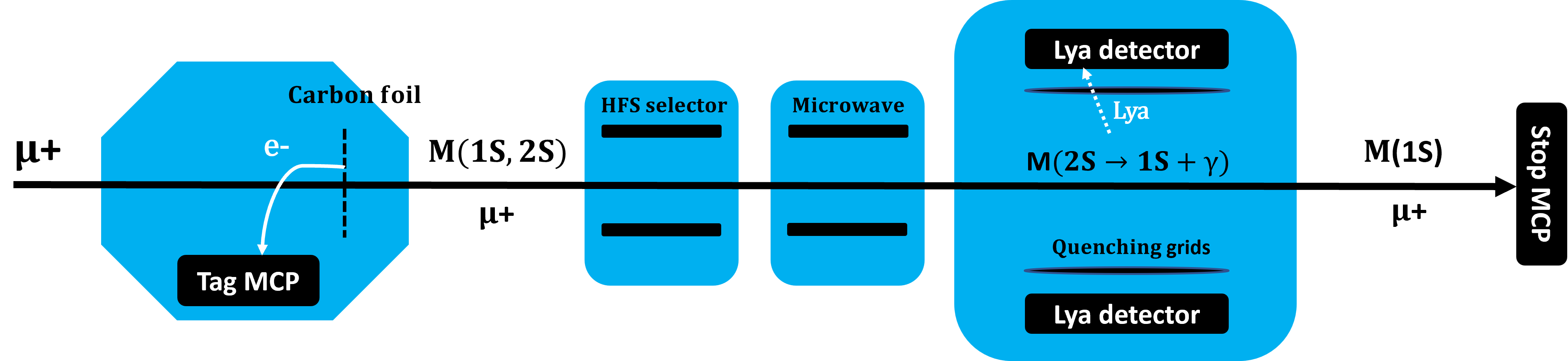}
    \caption{The experimental sketch of the muonium Lamb shift measurement within the Mu-MASS experiment at PSI.}
    \label{fig:lamb_shift_experimentl}
\end{figure}

The M(2S) passes through two microwave regions, one to depopulate unwanted hyperfine states and the other to drive the Lamb shift transition $2S$-$2P_{1/2}$ of interest. Once reaching the $2P$ state, the atom relaxes back to the ground state with a lifetime of \SI{1.6}{\nano\second}. The beam reaches then the detection chamber, where an electric field of the order of \SI{300}{\volt\per\cm} is quenching the remaining $2S$ atoms. The emitted Ly$\alpha$ photons (wavelength of \SI{122}{\nano\meter}) are detected by two coated MCP detectors. The muonium beam is eventually stopping on an MCP at the end of the beamline, creating the stop signal. The experimental sketch is shown in Fig.~\ref{fig:lamb_shift_experimentl}. By demanding a signal in all three detectors in specific time windows, the amount of Ly$\alpha$ photons detected at the applied microwave frequency can be extracted. At the frequency where the minimal amount of Ly$\alpha$ was detected, the microwave was optimally tuned and therefore depopulating the metastable beam most efficiently already before reaching the detection chamber. 

The M Lamb shift was determined to be \SI{1047.2+-2.5}{\mega\hertz} \cite{2021_Ohayon}, where the uncertainty is almost completely of statistical origin. This result is an improvement of an order of magnitude compared to the last best determinations \cite{1990_Woodle, 1984_Oram}. As a next step, the foil thickness will be reduced from \SI{10}{\nano\meter} to a few layers of graphene \cite{2014_Allegrini, 2014_Ebert} ($\sim$\SI{1}{\nano\meter}), which will increase our detectable M(2S) rate by a factor ranging from \numrange{15}{25}.

\section{Conclusions}
\label{sec-3}

Calculating each contribution separately shows that when reaching an uncertainty of around \SI{160}{\kilo\hertz} in muonium, one becomes sensitive to higher-order recoil effects such as E$_\text{BKG}$, which cannot be resolved yet with hydrogen. In the context of Mu-MASS, by replacing the \SI{10}{\nano\meter}-thick carbon foil for producing muonium with a few layers of graphene, an uncertainty of \SI{160}{\kilo\hertz} would be feasible within around a week of beamtime. Lowering the uncertainty further to \SI{40}{\kilo\hertz}, the effect coming from the nucleus self energy E$_\text{SEN}$ can be studied, which is also not in reach right now with hydrogen. The ongoing development of the muCool project \cite{2021_antognini} at PSI will improve the beam quality and provide an order of magnitude larger muon flux, allowing us to achieve such uncertainty. The realization of the High Intensity Muon Beam (HIMB) project at PSI \cite{Aiba:2021bxe} would allow to push this even further as summarized in Tab. \ref{tab:summary_improvement}. 

\begin{table}[b!h!t]

  \centering
  \caption[]{Summary of possible upgrades to the Mu-MASS setup and which contributions of the M LS could be probed.}
    \begin{tabular}{cccccc}
    \hline \noalign{\vspace{-1pt}} \hline
        \noalign{\smallskip}
     Beamline  & Target & Timeline & M(2S)& LS Uncertainty & Contributions \\
     & & & (\si{\hertz}) & (\si{\kilo\hertz}~/~\SI{10}{\day}) &  \\
        \noalign{\smallskip}
         \hline
        \noalign{\smallskip}

    PiE4/LEM & C-Foil & \num{2021} & \num{5} & \num{1000} & $E_\text{SE}$, $E_\text{VP}$, $E_\text{rec,S}$ \\
    PiE4/LEM & Graphene & \num{2023} & \num{100} & \num{200} &  $E_\text{BKG}$\\
    PiE1/muCool & Graphene & \num{2025} & \num{1000}& \num{70} & $E_\text{2ph}$ \\
    PiE1/muCool & Gas & \num{2026} & \num{5000} & \num{30} & $E_\text{SEN}$\\
    HiMB/muCool & Gas & \num{2029} & \num{100000} & \num{10} & $E_\text{RR}$, $E_\text{HFS}$, $E_\text{rec,R}$ \\ 
    \noalign{\smallskip}
    \hline \noalign{\smallskip} \hline
    \noalign{\vspace{2pt}}
\end{tabular}
\label{tab:summary_improvement}
\end{table}
\section*{Acknowledgements}
\label{acknowledgements}
This work is supported by the ERC consolidator grant 818053-Mu-MASS and the Swiss National Science Foundation under the grant 197346.
BO acknowledges support from the European Union’s Horizon 2020 research and innovation program under the Marie Skłodowska-Curie grant agreement No.~101019414. We are very grateful to K. Pachucki for his help with the theoretical calculations, S. Karshenboim, M. Eides and A. Czarnecki for the very useful discussions and T. Udem for sharing his code for the hydrogen calculations.

%
% BibTeX or Biber users please use (the style is already called in the class, ensure that the "woc.bst" style is in your local directory)
\bibliography{bibliography}

\begin{thebibliography}{33}

\bibitem{1981_Lundeen}
S.R. Lundeen, F.M. Pipkin, Phys. Rev. Lett. \textbf{46}, 232 (1981)

\bibitem{2014_CODATA}
P.J. Mohr, D.B. Newell, B.N. Taylor, Rev. Mod. Phys. \textbf{88}, 035009 (2016)

\bibitem{2010_Pohl}
R.~Pohl et~al., Nature \textbf{466}, 213 (2010)

\bibitem{2013_Antognini}
A.~Antognini, F.~Nez, K.~Schuhmann, F.D. Amaro, F.~Biraben, J.M.R. Cardoso,
  D.S. Covita, A.~Dax, S.~Dhawan, M.~Diepold et~al., Science \textbf{339}, 417
  (2013)

\bibitem{2013_Pohl}
R.~Pohl, R.~Gilman, G.A. Miller, K.~Pachucki, Annual Review of Nuclear and
  Particle Science \textbf{63}, 175 (2013)

\bibitem{2019_Xiong}
W.~Xiong, A.~Gasparian, H.~Gao, D.~Dutta, M.~Khandaker, N.~Liyanage, E.~Pasyuk,
  C.~Peng, X.~Bai, L.~Ye et~al., Nature (London) \textbf{575} (2019)

\bibitem{2019_Hessels}
N.~Bezginov, T.~Valdez, M.~Horbatsch, A.~Marsman, A.C. Vutha, E.A. Hessels,
  Science \textbf{365}, 1007 (2019)

\bibitem{2017_Beyer}
A.~Beyer, L.~Maisenbacher, A.~Matveev, R.~Pohl, K.~Khabarova, A.~Grinin,
  T.~Lamour, D.C. Yost, T.W. Hänsch, N.~Kolachevsky et~al., Science
  \textbf{358}, 79 (2017)

\bibitem{2021_Brandt}
A.D. Brandt, S.F. Cooper, C.~Rasor, Z.~Burkley, D.C. Yost, A.~Matveev,
  \emph{Measurement of the $2$s$_{1/2}-8$d$_{5/2}$ transition in hydrogen}
  (2021), \texttt{2111.08554}

\bibitem{2020_Grinin}
A.~Grinin, A.~Matveev, D.C. Yost, L.~Maisenbacher, V.~Wirthl, R.~Pohl, T.W.
  Hänsch, T.~Udem, Science \textbf{370}, 1061 (2020)

\bibitem{2018_Fleurbaey}
H.~Fleurbaey, S.~Galtier, S.~Thomas, M.~Bonnaud, L.~Julien, F.m.c. Biraben,
  F.m.c. Nez, M.~Abgrall, J.~Gu\'ena, Phys. Rev. Lett. \textbf{120}, 183001
  (2018)

\bibitem{2020_Karr}
J.P. Karr, D.~Marchand, E.~Voutier, Nature Reviews Physics pp. 1--14 (2020)

\bibitem{1990_Sapirstein}
J.R. Sapirstein, D.R. Yennie, Adv. Ser. Direct. High Energy Phys. \textbf{7},
  560 (1990)

\bibitem{2021_CODATA}
E.~Tiesinga, P.J. Mohr, D.B. Newell, B.N. Taylor, Journal of Physical and
  Chemical Reference Data \textbf{50}, 033105 (2021)

\bibitem{1995_Pachucki}
K.~Pachucki, Phys. Rev. A \textbf{52}, 1079 (1995)

\bibitem{2019_Pachucki}
V.A. Yerokhin, K.~Pachucki, V.~Patkóš, Annalen der Physik \textbf{531},
  1800324 (2019)

\bibitem{2019-Saveli}
S.G. Karshenboim, A.~Ozawa, V.G. Ivanov, Phys. Rev. A \textbf{100}, 032515
  (2019)

\bibitem{2019-Saveli2}
S.G. Karshenboim, A.~Ozawa, V.A. Shelyuto, R.~Szafron, V.G. Ivanov, Physics
  Letters B \textbf{795}, 432 (2019)

\bibitem{2002-Czar}
I.~Blokland, A.~Czarnecki, K.~Melnikov, Phys. Rev. D \textbf{65}, 073015 (2002)

\bibitem{1995-Eides}
M.I. Eides, H.~Grotch, Phys. Rev. A \textbf{52}, 1757 (1995)

\bibitem{2021-Eides}
M.I. Eides, V.A. Shelyuto, \emph{Three-loop corrections to lamb shift in
  muonium and positronium} (2021), \texttt{2110.13279}

\bibitem{2019_Peset}
C.~Frugiuele, J.~P\'erez-R\'{\i}os, C.~Peset, Phys. Rev. D \textbf{100}, 015010
  (2019)

\bibitem{2018_Crivelli}
P.~Crivelli, Hyperfine Interactions \textbf{239} (2018)

\bibitem{2012_Antognini}
A.~Antognini, P.~Crivelli, T.~Prokscha, K.S. Khaw, B.~Barbiellini, L.~Liszkay,
  K.~Kirch, K.~Kwuida, E.~Morenzoni, F.M. Piegsa et~al., Phys. Rev. Lett.
  \textbf{108}, 143401 (2012)

\bibitem{2021_Burkley}
Z.~Burkley, L.~de~Sousa~Borges, B.~Ohayon, A.~Golovozin, J.~Zhang, P.~Crivelli,
  Opt. Express \textbf{29}, 27450 (2021)

\bibitem{2020_Janka}
G.~Janka, B.~Ohayon, Z.~Burkley, L.~Gerchow, N.~Kuroda, X.~Ni, R.~Nishi,
  Z.~Salman, A.~Suter, M.~Tuzi et~al., Eur. Phys. J. C \textbf{80}, 804 (2020)

\bibitem{2021_Ohayon}
B.~Ohayon, G.~Janka, I.~Cortinovis, Z.~Burkley, L.~de~Sousa~Bourges, E.~Depero,
  A.~Golovizin, X.~Ni, Z.~Salman, A.~Suter et~al., \emph{Precision measurement
  of the lamb shift in muonium} (2021), \texttt{2108.12891}

\bibitem{1990_Woodle}
K.A. Woodle, A.~Badertscher, V.W. Hughes, D.C. Lu, M.W. Ritter, M.~Gladisch,
  H.~Orth, G.~zu~Putlitz, M.~Eckhause, J.~Kane et~al., Phys. Rev. A
  \textbf{41}, 93 (1990)

\bibitem{1984_Oram}
C.J. Oram, J.M. Bailey, P.W. Schmor, C.A. Fry, R.F. Kiefl, J.B. Warren, G.M.
  Marshall, A.~Olin, Phys. Rev. Lett. \textbf{52}, 910 (1984)

\bibitem{2014_Allegrini}
F.~Allegrini, R.W. Ebert, S.A. Fuselier, G.~Nicolaou, P.V. Bedworth, S.W.
  Sinton, K.J. Trattner, Optical Engineering \textbf{53}, 1  (2014)

\bibitem{2014_Ebert}
R.W. Ebert, F.~Allegrini, S.A. Fuselier, G.~Nicolaou, P.~Bedworth, S.~Sinton,
  K.J. Trattner, Review of Scientific Instruments \textbf{85}, 033302 (2014)

\bibitem{2021_antognini}
A.~Antognini, D.~Taqqu, SciPost Phys. Proc. \textbf{5}, 030 (2021)

\bibitem{Aiba:2021bxe}
M.~Aiba et~al. (2021), \texttt{2111.05788}

\end{thebibliography}

\end{document}